\documentclass{elsart}
\usepackage[dvips]{graphicx}

\usepackage{color}
\usepackage{amssymb}
\usepackage{amsmath}
\usepackage{epsf}
\usepackage{subfigure}
\newcommand{\rmd}{\text{d}}
\newcommand{\dfracp}[2]{\dfrac{\partial #1}{\partial #2}}
\newcommand{\dfracd}[2]{\dfrac{\text{d} #1}{\text{d} #2}}

\begin{document}

\begin{frontmatter}
\title{Apparent fractal dimensions in the HMF model}

\author{Luca Sguanci$^1$}, \author{Dieter H. E. Gross$^2$}
and \author{Stefano Ruffo$^{3}$}

\address{1. Dipartimento di Energetica ``S. Stecco", 
Universit{\`a} di Firenze, Via S.Marta,
3 I-50139, Firenze, Italy} 

\address{2. Hahn Meitner Institute, Bereich SF5
Glienicker Str.100, D14109 Berlin, Germany} 

\address{3. Dipartimento di Energetica
``S. Stecco" and CSDC, Universit{\`a} di Firenze, 
INFM and INFN, Via S.Marta,
3 I-50139, Firenze, Italy}

\thanks[luca]{E-mail: luke@dma.unifi.it}
\thanks[dieter]{E-mail: gross@hmi.de}
\thanks[stefano]{E-mail: ruffo@avanzi.de.unifi.it}

\begin{abstract}
We show that recent observations of fractal dimensions in the $\mu$-space
of $N$-body Hamiltonian systems with
long-range interactions are due to finite $N$ and finite resolution effects.
We provide strong numerical evidence that, in the continuum (Vlasov) limit,
a set which initially is not a fractal
(e.g. a line in 2D) remains such for all finite times. We perform this
analysis for the Hamiltonian Mean Field (HMF) model, which describes the
motion of a system of $N$ fully coupled rotors. The analysis can be
indirectly confirmed by studying the evolution of a large set of initial points
for the Chirikov standard map.
\end{abstract}
\end{frontmatter}

It has been recently claimed~\cite{hiroko,tsallis} that structures
characterized by a non integer fractal dimension~\cite{falconer}
form spontaneously in the $\mu$-space of some $N$-body Hamiltonian systems,
even when starting from a set of initial conditions (points
in $\mu$-space) that does not show any fractality. 

The model analysed in Ref.~\cite{hiroko}
is the one-dimensional self-gravitating system~\cite{hohlfeix} and the
initial condition is a {\it line} in $\mu$-space, corresponding to a vanishing
virial ratio (zero velocity for all mass sheets).
The model studied in Ref.~\cite{tsallis} describes the motion of
a fully coupled system of $N$ rotors and goes
under the name of Hamiltonian Mean Field (HMF) 
model~\cite{antoniruffo}; the initial
condition is such that all rotors are fixed at some given angle (conventionally
zero) and angular velocities are uniformly spread in an
interval, symmetric around zero. 

There is a strong analogy
between the two results since: {\it i)} both models possess a 
continuum mean-field limit ($N\to\infty$ at fixed volume)
for which the single particle distribution function (SPDF)
$f(q,p,t)$ evolves according to a Vlasov-Poisson system~\cite{Spohn},
{\it ii)} both initial states are {\it lines} in $\mu$-space 
when the continuum limit is taken.

In this short note we analyse the HMF model, considering the same initial
state as the one studied in Ref.~\cite{tsallis}. We give compelling
numerical evidence that, as the number of rotors grows
and as the resolution increases, the collection of initial points 
lies on a set of dimension one in the $\mu$-space (which is two-dimensional
for this model) at all finite times. 

We will explain that this is a somewhat trivial consequence of the
fact that the SPDF obeys a Vlasov equation in the mean-field limit.
Moreover, we will show that {\it apparent fractal dimensions} can be indeed 
observed at low resolutions, justifying the observations of Ref.~\cite{tsallis}.
For comparison, we will show what happens for the standard Chirikov's
map~\cite{chirikov}, which shares the symplectic property with the HMF
model, for which we can push the analysis to a larger number of initial
points.

The Hamiltonian of the HMF model is
\begin{equation}
\label{HMF}
H_{HMF} = \dfrac{1}{2} \sum_{j=1}^{N} p_{j}^{2} + 
\dfrac{1}{2N} \sum_{j,k=1}^{N} [ 1-\cos(q_{j}-q_{k}) ],
\end{equation}
where $q_i\in[-\pi,\pi[$ is the position (angle) of the
$i$-th particle on a circle and $p_i$ the corresponding conjugate
variable.
This system can be seen as representing particles moving on a unit
circle interacting via an infinite range attractive cosine
potential, or as classical XY rotors with infinite range
ferromagnetic couplings (for more details see Ref.~\cite{HMFSpringer}).
The magnetization,defined as
\begin{equation}
\label{magnetization}
\overrightarrow{M}(t) = (M_x,M_y) = \dfrac{1}{N} \sum_{j=1}^{N}
( \cos q_{j}(t),\sin q_{j}(t)),
\end{equation}
is the main observable that characterizes the dynamical and 
thermodynamic state of the system.

In the continuum limit, that is keeping the volume (here the
interval $[-\pi,\pi[$) and the energy per particle fixed as the
number of particles $N\to \infty$, the dynamics governed by
Hamiltonian (\ref{HMF}) is described by a Vlasov equation. 
Indeed, the state of the finite $N$ system can be described by a single
particle time-dependent distribution function
\begin{equation}
f_d\left(q,p,t\right)= \displaystyle \frac{1}{N} \sum_{j=1}^N\delta
\left(q -q_{j}\left( t\right) ,p-p_{j}\left( t\right) \right)~,
\end{equation}
where $\delta$ is the Dirac function. When $N$ is large, it is
natural to approximate the discrete density $f_d$ by a continuous
SPDF $f\left(q,p,t\right)$. Using this distribution, one can 
rewrite the two components of the magnetization~$M$ as 
\begin{eqnarray}
\label{magnetization_Vlasov}
\overline{M}_x\left[ f\right] &\equiv&
\int f(q,p,t)\, \cos q\ \rmd q \rmd p\quad,\\
\overline{M}_y\left[ f\right] &\equiv &
\int f(q,p,t)\,\sin q \ \rmd q \rmd p\quad.
\label{magnetization_Vlasovb}
\end{eqnarray}
Within this approximation the potential that affects all the particles is
\begin{equation}
V\left(q \right) \left[ f\right] =
1-\overline{M}_x\left[ f\right]\, \cos q -
\overline{M}_y\left[ f\right]\, \sin q \quad. 
\end{equation}
This potential enters the expression of the Vlasov equation
\begin{equation}
\label{vlasov}
\dfracp {f}{t} + p \dfracp {f}{q}
-\dfracd{V}{q} \left[ f\right] \dfracp{f}{p} = 0\quad,
\end{equation}
which governs the spatio-temporal evolution of the SPDF $f$.

The initial condition considered in ~\cite{tsallis} is
$q_i=0,\forall i$ and $p_i$ random i.i.d. uniformly distributed
in the interval $[-\bar{p},\bar{p}]$. As usual, we call
this class of initial conditions {\it water bags} (WB). The value of $\bar{p}$
is chosen such that the energy per particle is $U=H_{HMF}/N=0.69$
and $N=10000$, but these data will be varied in this paper. 
For such an energy this initial state is 
Vlasov unstable~\cite{Yama}, and hence the SPDF evolves in time
(see fig. 3 in Ref.~\cite{tsallis}).

However, since initially the state is a {\it line} in the
two-dimensional $\mu$-space $(q,p)$, it will remain 
a line for all finite times, although intricately stretched and folded.
This is a trivial property related to the fact that Vlasov
equation defines a {\it flow} $\Phi_t$ for the particles (points ${\bf x}=(q,p)$ 
in the $\mu$-space continuum)
\begin{equation}
{\bf x}(t)=\Phi_t({\bf x}(0))~,
\end{equation}
which is a diffeomorfism. Moreover, since the Jacobian 
$J=\partial {\bf x}(t)/\partial {\bf x}(0)$ is a symplectic
matrix, areas modify their shape but are conserved.

The reason why the set does not appear to be a line is due 
to the finite particle number in the simulation. The line
is stretched and folded by the two main phase-space mechanisms: the formation 
of whorls around elliptic fixed points and of tendrils in 
the homoclinic tangle~\cite{ottino}. Moreover, local hyperbolicity
in phase-space produces a dispersion of the initial points. 
As a consequence, the set of initial points, appears by visual
inspection, in the course of time, to be partly distributed
along lines and partly dispersed onto the two-dimensional $\mu$-space
(see again fig. 3 in Ref.~\cite{tsallis} or fig. 2 in Ref.~\cite{hiroko})
This is why the authors of Refs.~\cite{hiroko,tsallis} have used 
{\it fractal dimension} to chacterize this set, more specifically
box dimension~\cite{hiroko} and correlation dimension~\cite{tsallis}.

\begin{figure}[htbp]
  \centering
  \includegraphics[width=7cm]{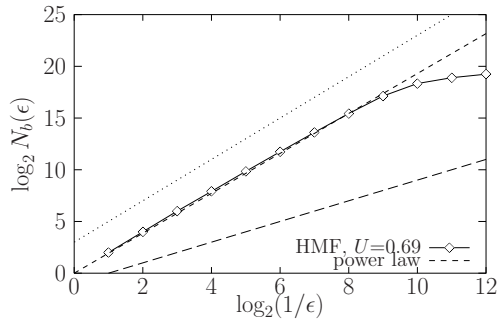}
  \caption{Number of occupied boxes of linear size $\epsilon$ versus
  resolution $1/\epsilon$ in log-log scale for the HMF model in a WB initial 
  condition with $U=0.69$ and $N=10^6$ ($\Diamond$). The short-dashed
  line superposed to the data is a power law fit with {\it apparent fractal dimension}
  $D_0 \sim 1.93$. The upper deviation from the power law is at
  $1/\epsilon=2^{10}$, i.e. roughly when the number of boxes $(1/\epsilon)^2$
  is of the order of the number of points $N$. The dotted and long-dashed
  lines are the power laws with exponents $2$ and $1$, respectively.}
  \label{fig1}
\end{figure}

The number of boxes $N_b(\epsilon)$ of linear size $\epsilon$ 
which cover the set of $N=10^6$ points at time $t=75$
when starting the HMF model in a WB initial condition at $U=0.69$
is shown as a function of resolution $1/\epsilon$ in log-log scale
in fig.~\ref{fig1}.
These data are not qualitatively different from the ones in fig.~3 of
Ref.~\cite{hiroko} and one might then fit the central part of them with
the Ansatz $N_b=\epsilon^{-D_0}$ (since the maximal box size which
contains all the set is normalized to one), obtaining $D_0 \sim 1.930 \pm 0.002$ 
and concluding that the set is a fractal, with dimension $D_0$ close 
to two, but definitely distinct from two within numerical accuracy. This
is a consequence of the set being a mixture of a set with one-dimensional
features (the initial line) and two-dimensional ones caused by chaotic spread.

\begin{figure}[htbp]
  \centering
        \includegraphics[width=7cm]{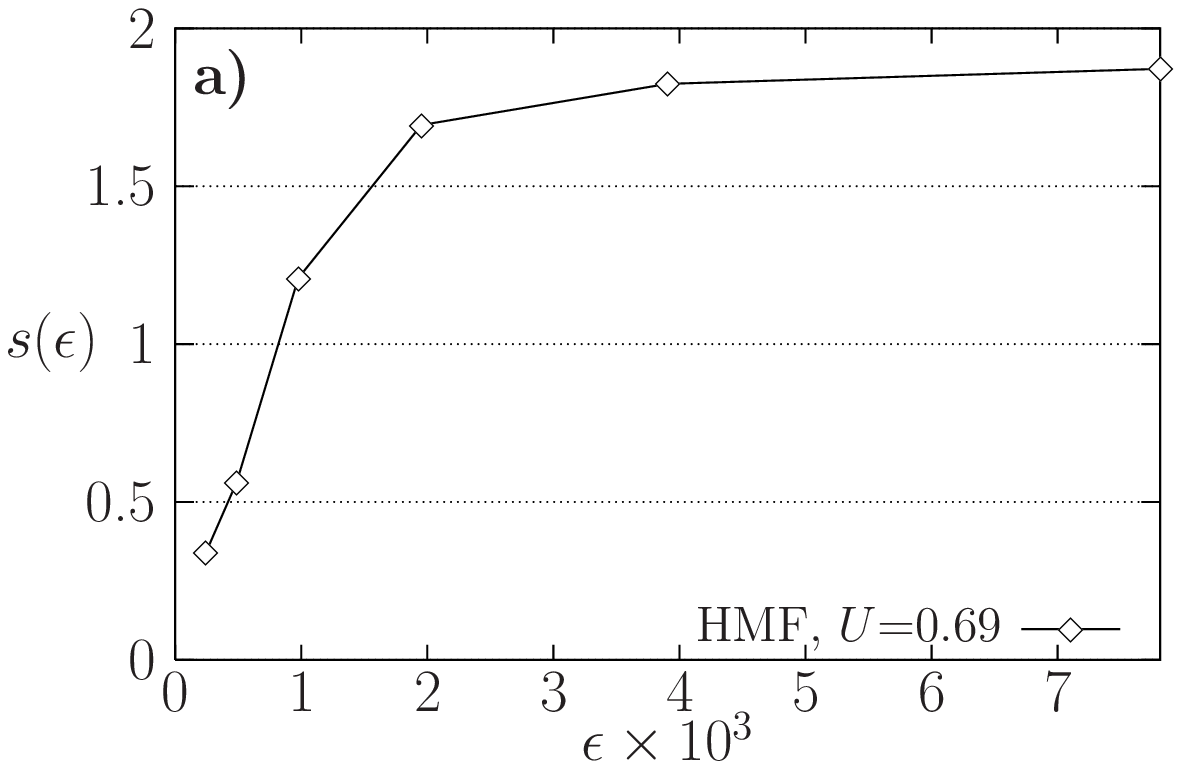}
	\includegraphics[width=7cm]{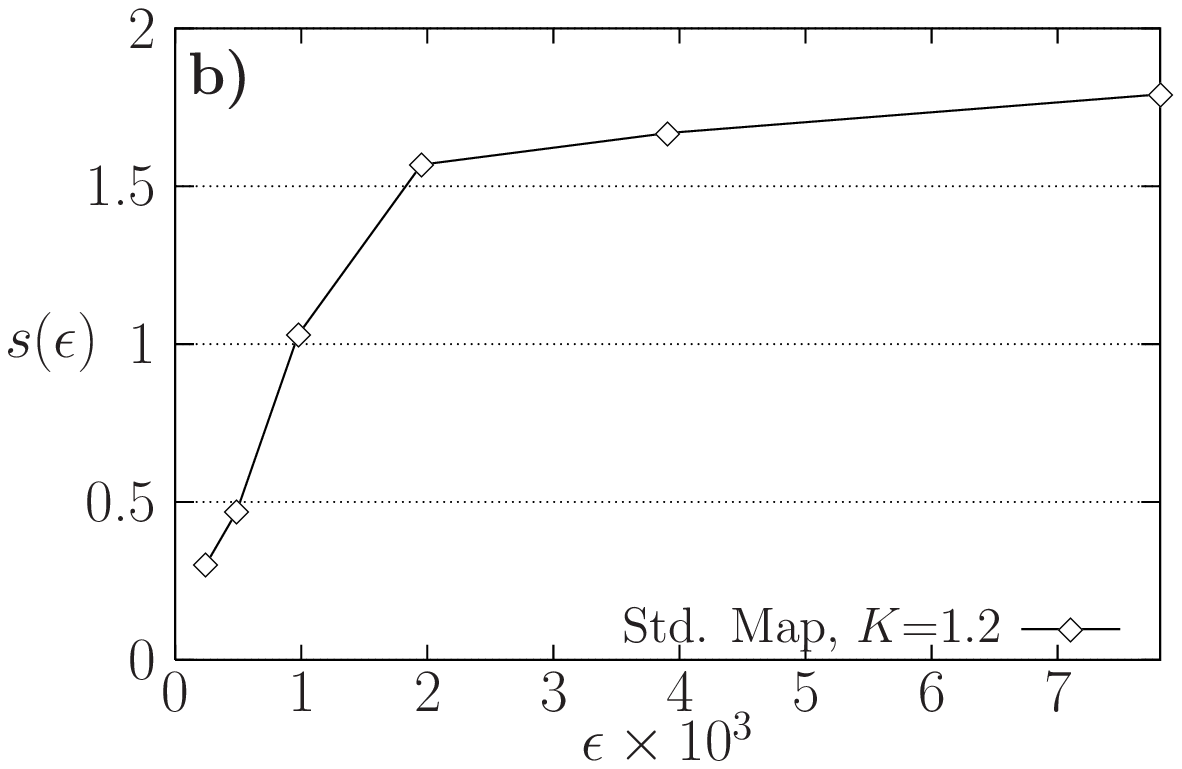}
	\caption{Local slopes $s(\epsilon)$ for resolutions 
	ranging from $1/\epsilon=2^7$ up to $2^{12}$. In Figure~{\bf a)} we plot the results 
	for the HMF model at $t=75$ with WB initial condition at $U=0.69$ and $N=10^6$. 
	Figure~{\bf b)} refers to the standard map at $t=160$ for $K=1.2$ and 
	$NP=10^6$.}
	\label{fig2}
\end{figure}

\begin{figure}[htbp]
  \centering
  \includegraphics[width=7cm]{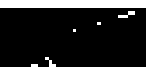}
  \includegraphics[width=7cm]{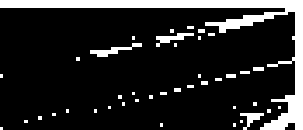}
  \includegraphics[width=7cm]{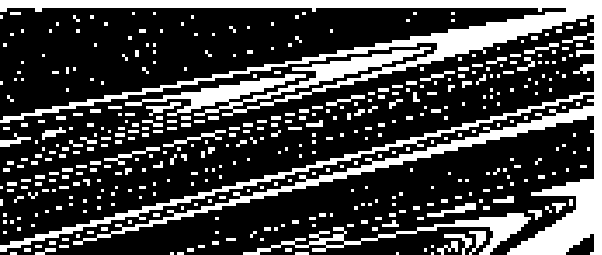}
  \includegraphics[width=7cm]{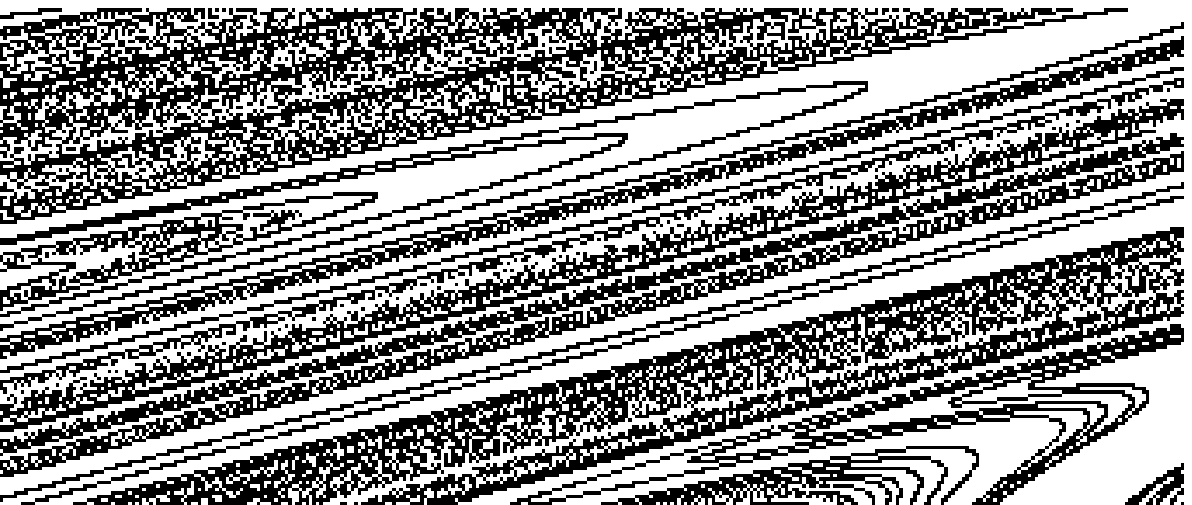}
  \includegraphics[width=7cm]{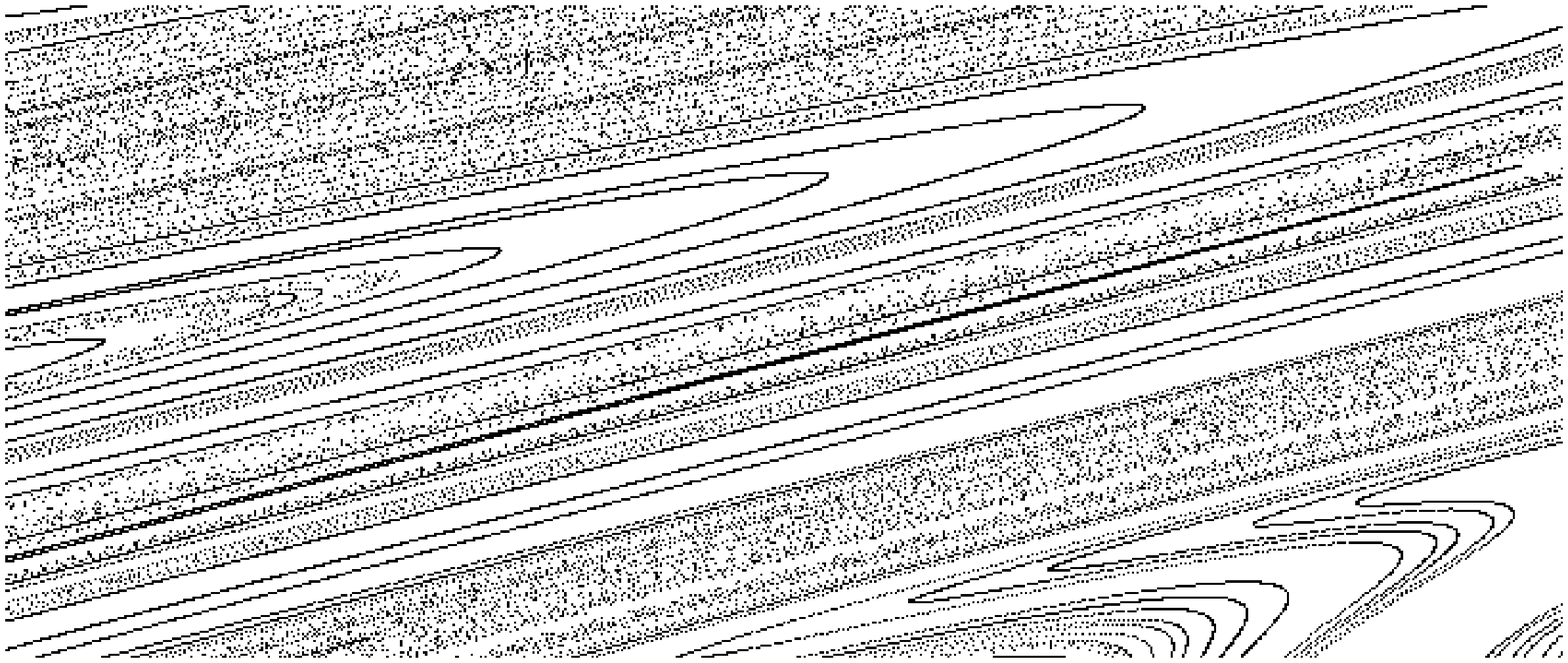}
  \includegraphics[width=7cm]{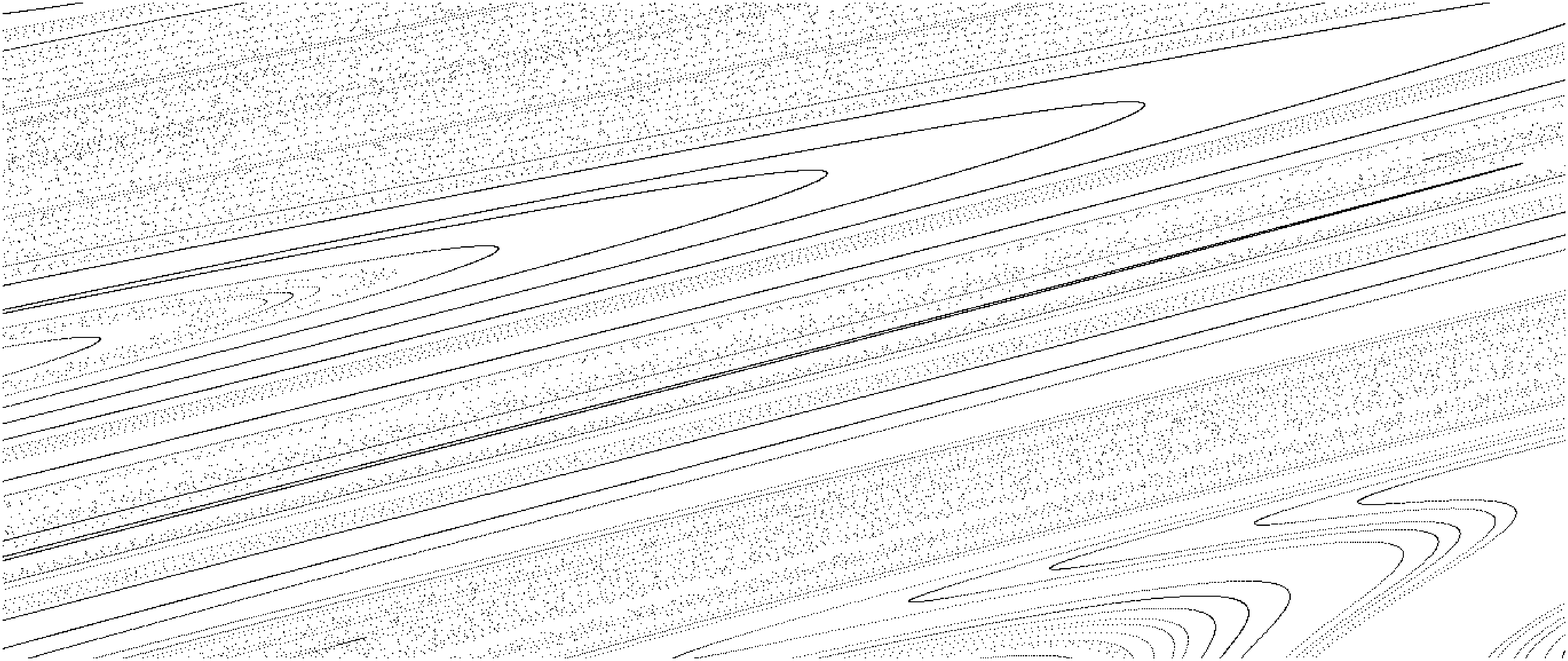}
	\caption{HMF model. A set of $N=10^{6}$ points is, initially, uniformly
	distributed along the $p$ axis between $-\bar{p}$ and $\bar{p}$, such
	that $U=\bar{p}^2/6=0.69$. From left to right and from top to bottom 
	we show a portion $S=[0.33,0.66[\times[0.36,0.50[$ of the $\mu$-space,
	where a box of given linear dimension $\epsilon$ is colored black if
	at least one point of the initial condition is contained inside it
	at time $t=75$. Resolution ranges from $1/\epsilon=2^{7}$ up to 
	$1/\epsilon=2^{12}$.}
	\label{fig3}
\end{figure}

However, as first remarked in Ref.~\cite{benettin} the slope of the
curve in fig.~\ref{fig1} is not constant and depends on the resolution.
It is already evident in this figure that, as the resolution increases
($\epsilon \to 0$), the local slope tends to zero (on this scale most of the points
are surrounded by very few neighbours). This is better shown in 
fig.~\ref{fig2}a, where the discrete approximation of the local slope in logarithmic scale 
\begin{equation}
s(\epsilon)=\frac{d \log_2 N_b(\epsilon)}{d \log_2 (1/\epsilon)}~,
\label{local}
\end{equation}
is plotted as a function of $\epsilon$ for the same set of points. Going
from low to high resolution, all dimensions from 2 to 0 are covered continuously
and smoothly. All are {\it apparent fractal dimensions} of a set which
has indeed dimension 1. It might look strange that one does not observe
any signal of dimension 1 at any resolution. This is due to the
{\it inhomogeneity}, as shown in fig.~\ref{fig3}, where one
represents in black the occupied boxes at increasing resolutions. While
at low resolutions all boxes are occupied indicating a set of dimension 2,
when the resolution increases, the set concentrates on a smaller set of
boxes, but already at resolution $1/\epsilon=2^{9}$ one observes structures
of dimension 1 (lines). However, their weight in the $\mu$-space is negligible
and, as the resolution is increased even more, the fraction of the set made
by spread points dominates and consequently dimension 0. A complete
multifractal~\cite{falconer} analysis should be realized to highlight
such inhomogeneities.

\begin{figure}[htbp]
  \centering
	\includegraphics[width=7cm]{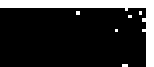}
	\includegraphics[width=7cm]{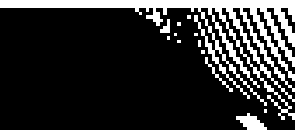}
	\includegraphics[width=7cm]{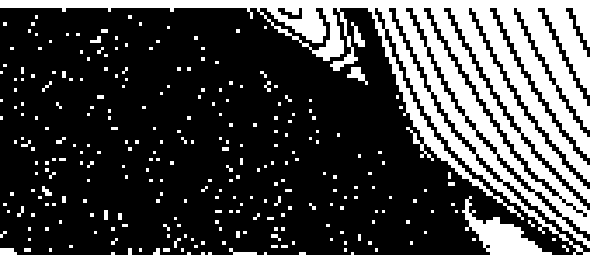}
	\includegraphics[width=7cm]{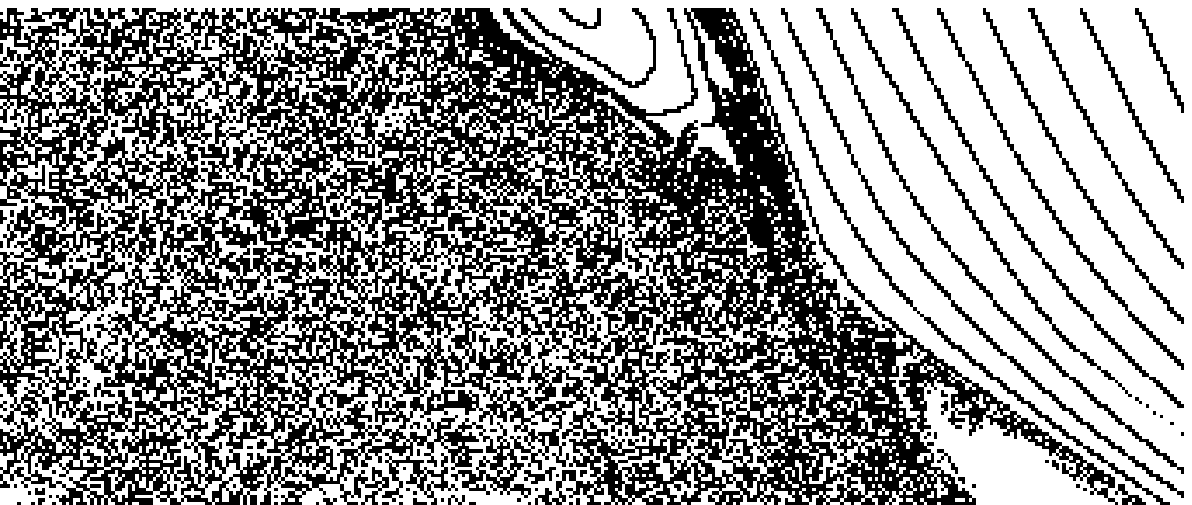}
	\includegraphics[width=7cm]{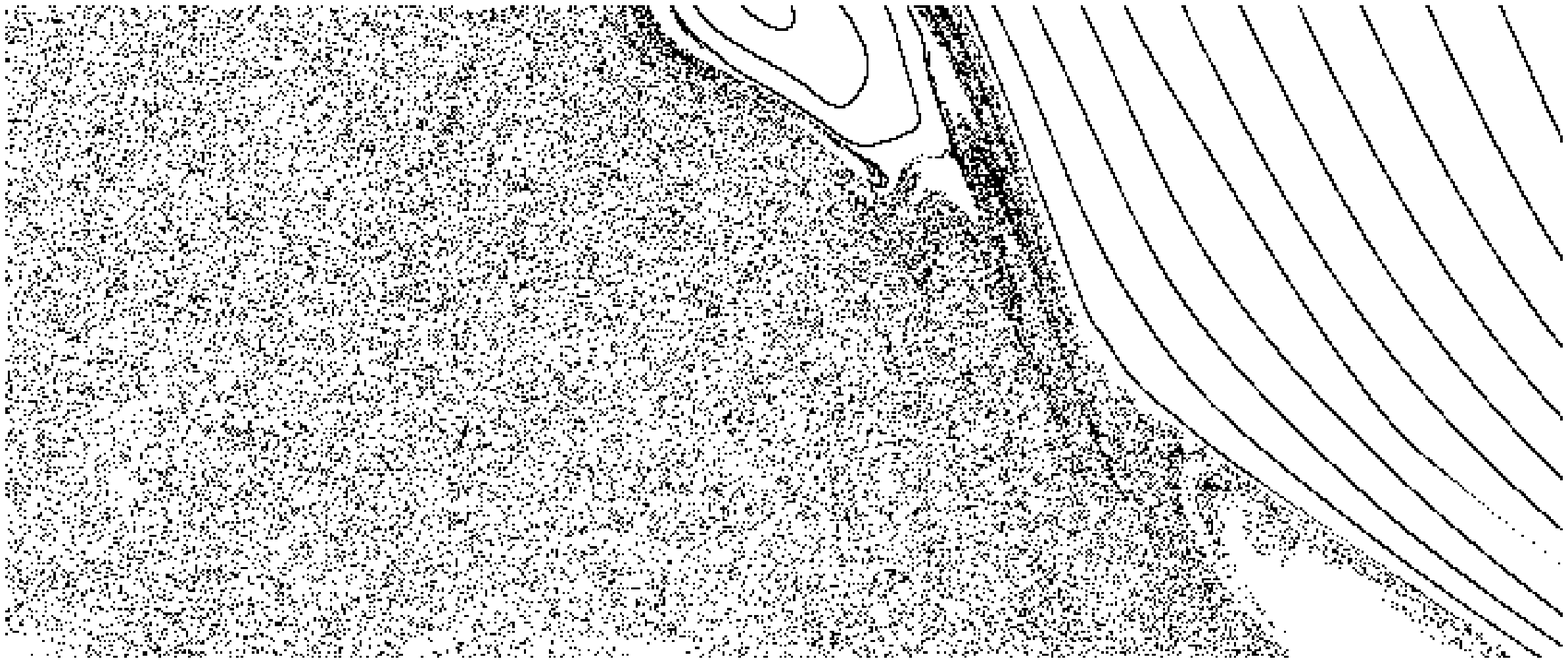}
	\includegraphics[width=7cm]{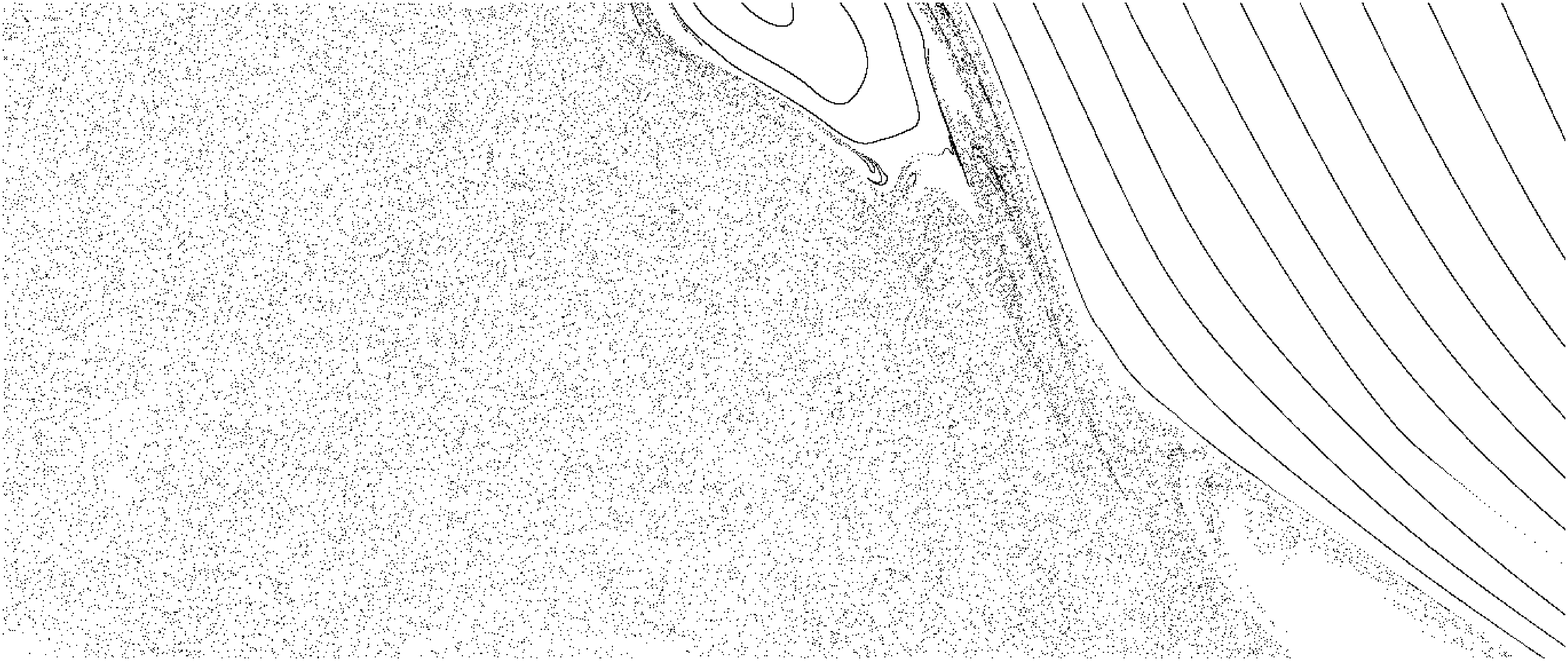}
	\caption{Standard map at $K=1.2$. A set of $NP=10^{6}$ particles 
	is, initially, uniformly
	distributed along the $p$ axis. From left to right and from top to bottom 
	we show a portion $S=[0,0.33[\times[0.86,1.00[$ of the phase space,
	where a box of given linear dimension $\epsilon$ is colored black if
	at least one particle of the initial condition is contained inside it
	at time $t=160$. Resolution ranges from $1/\epsilon=2^{7}$ up to 
	$1/\epsilon=2^{12}$.}
	\label{fig4}
\end{figure}

For comparison, we have performed a similar experiment for the Chirikov
standard map~\cite{chirikov} 
\begin{eqnarray}
p_{t+1}^i&=&p_t^i+q_t^i\\
q_{t+1}^i&=&q_t^i-K \sin (p_{t+1}^i) \quad \mbox{mod $2\pi$}~,
\label{stmap}
\end{eqnarray}
where the index $t=0,\ldots$ denotes time and the index $i=1,\ldots,NP$
the number of initial particles. Particles are put at $t=0$, as above,
at $q=0$ and with momentum uniformly spread in the interval $[-\bar{p},\bar{p}]$.
The phase space is closed by periodizing in $2\pi$  both the $q$ and the
$p$ directions and normalized in the unit square. 
The results of numerical experiments conducted above the
chaos threshold, at $K=1.2$, i.e. a parameter region with a {\it mixed}
(chaotic and regular) phase space, are shown in fig.~\ref{fig2}b (local slopes) and
fig.~\ref{fig4} (occupied boxes in phase space at different resolutions). 
The behavior of all measured quantities is qualitatively
the same as the one of the HMF model.

\begin{figure}[htbp]
  \centering
        \includegraphics[width=7cm]{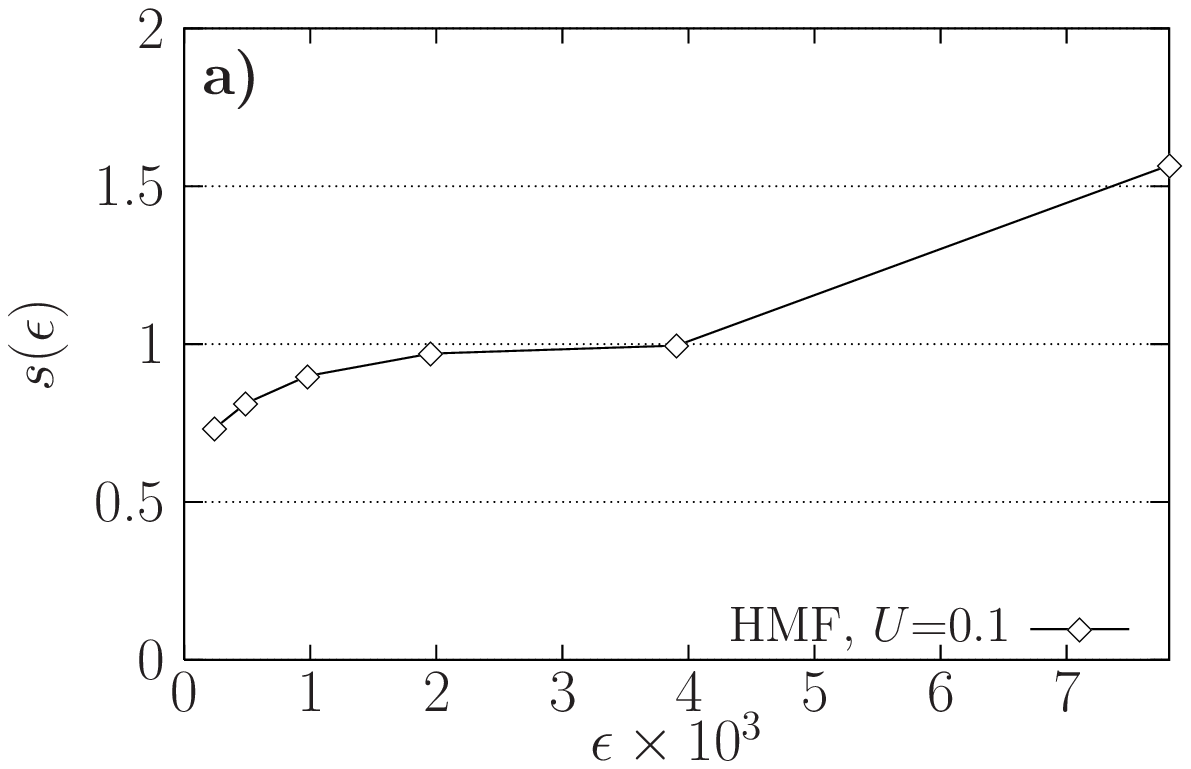}
	\includegraphics[width=7cm]{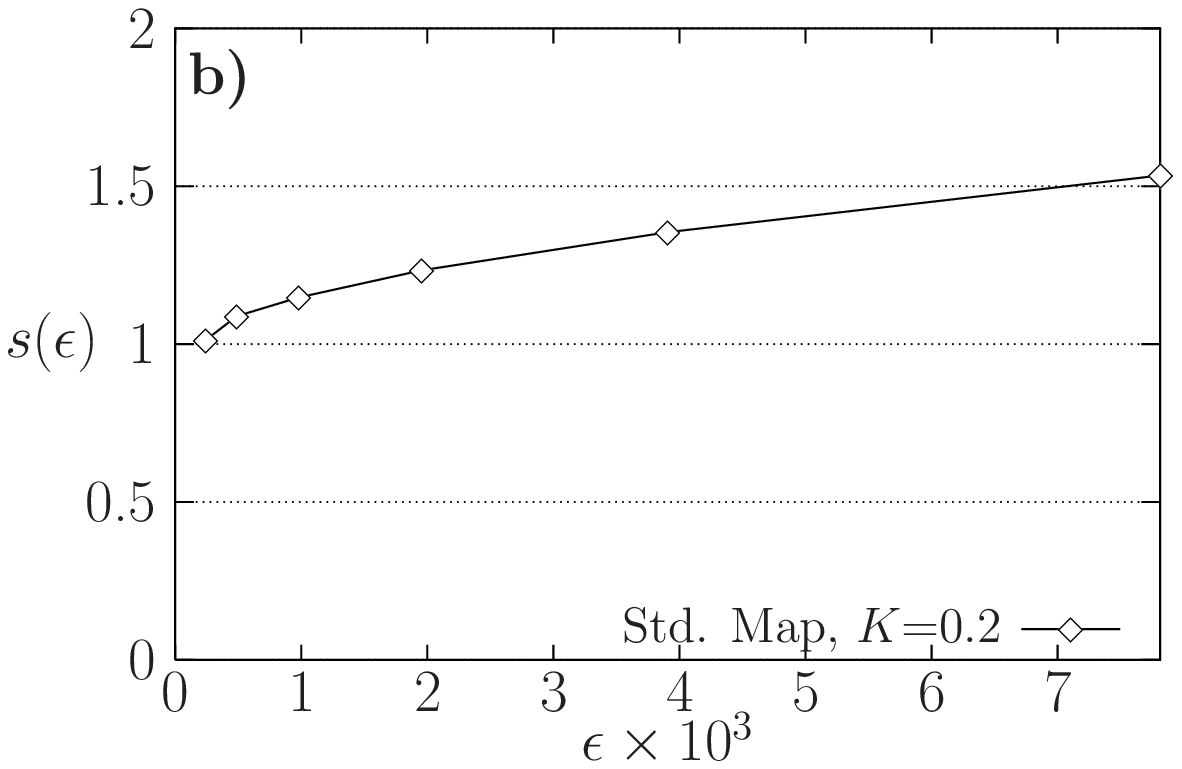}
	\caption{This figure shows how the local slope changes for resolutions 
	ranging from $1/\epsilon=2^8$ up to $2^{12}$. In Figure~{\bf a)} we plot the results 
	at $t=500$ for the HMF model with WB initial condition corresponding to 
	$U=0.1$ and a number of particles $N=10^6$.  Figure~{\bf b)} refers to 
	the Standard Map at $t=160$ for $K=0.2$ and $NP=10^6$.
	Again the ``natural" slope $s=1$ is roughly at $(1/\epsilon)^2\sim N,NP$.
	}
	\label{fig5}
\end{figure}

Can one get from straight box-counting an indication of the presence of sets
of dimension 1? To obtain this results it is enough to perform experiments
at lower energies for the HMF model and at smaller values of parameter
$K$ in the standard map. By doing this, one decreases the fraction of phase space
which is occupied by chaotic orbits and, hence, reduces particle dispersion.
In fig.~\ref{fig5} we plot the local slopes for the HMF model at $U=0.1$
(fig.~\ref{fig5}a) and at $K=0.2$ for the standard map (fig.~\ref{fig5}b). In these cases,
the local slope, and hence fractal dimension, shows an evident plateau
at the value $1$ as the resolution increases, before dropping to zero when
the individual points of the set are detected.
Let us remark that the number of particles is the same as before $N=NP=10^6$.
This is a strong evidence that the line remains a line during time 
evolution, as expected.

\begin{figure}[htbp]
  \centering
        \includegraphics[width=7cm]{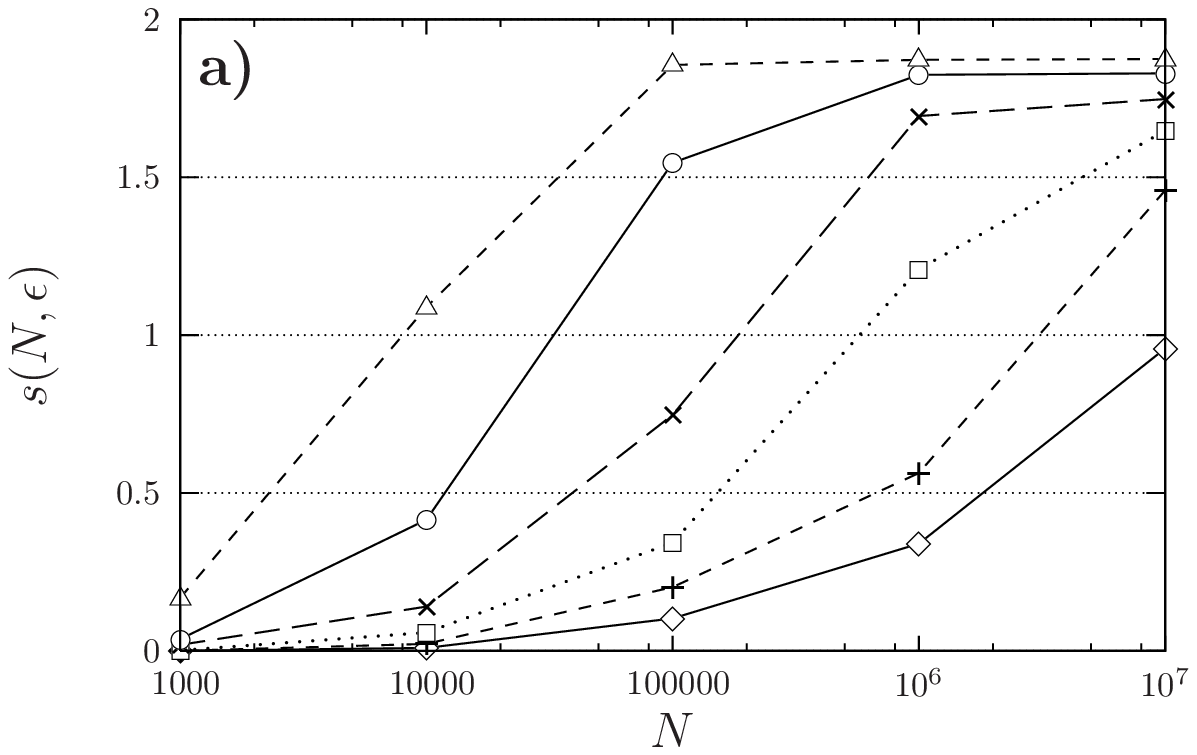}
	\includegraphics[width=7cm]{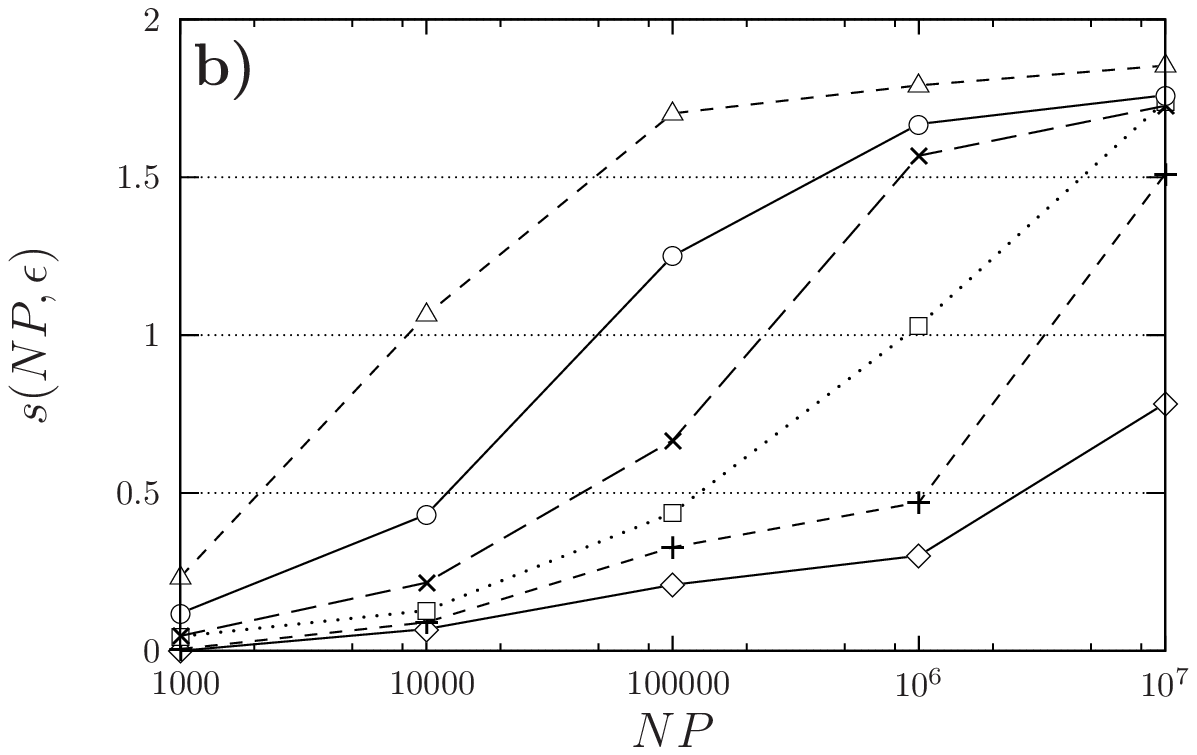}
	\caption{Local slope as a function of the number of particles 
	(ranging from $10^3$ up to $10^7$) for different values of the resolution 
	$1/\epsilon$: $2^{7}$ ($\triangle$); $2^{8}$($\circ$); $2^{9}$ ($\times$);
	$2^{10}$ ($\Box$); $2^{11}$ ($+$); $2^{12}$($\Diamond$).
	Figure~{\bf a)} shows the results for the HMF model at $U=0.69$ 
	and Figure~{\bf b)} the corresponding results for the standard map 
	at $K=1.2$.}
	\label{fig6}
\end{figure}

\begin{figure}[htbp]
  \centering
  \includegraphics[width=7cm]{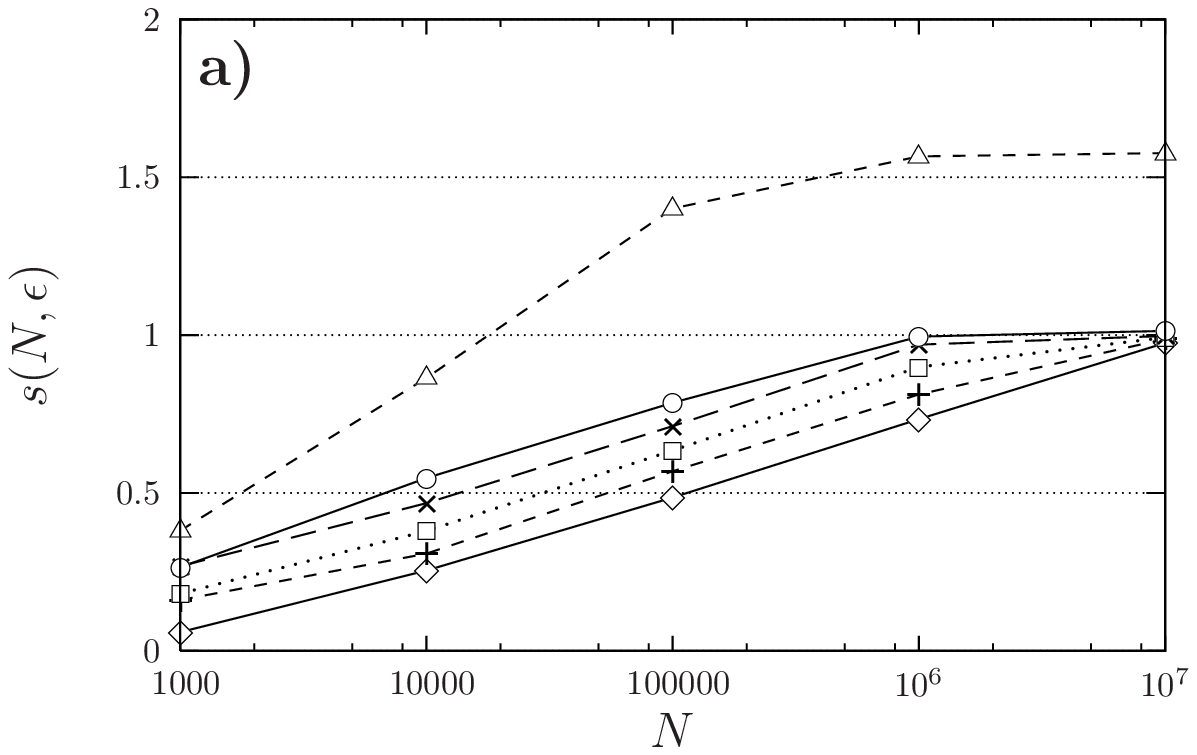}
	\includegraphics[width=7cm]{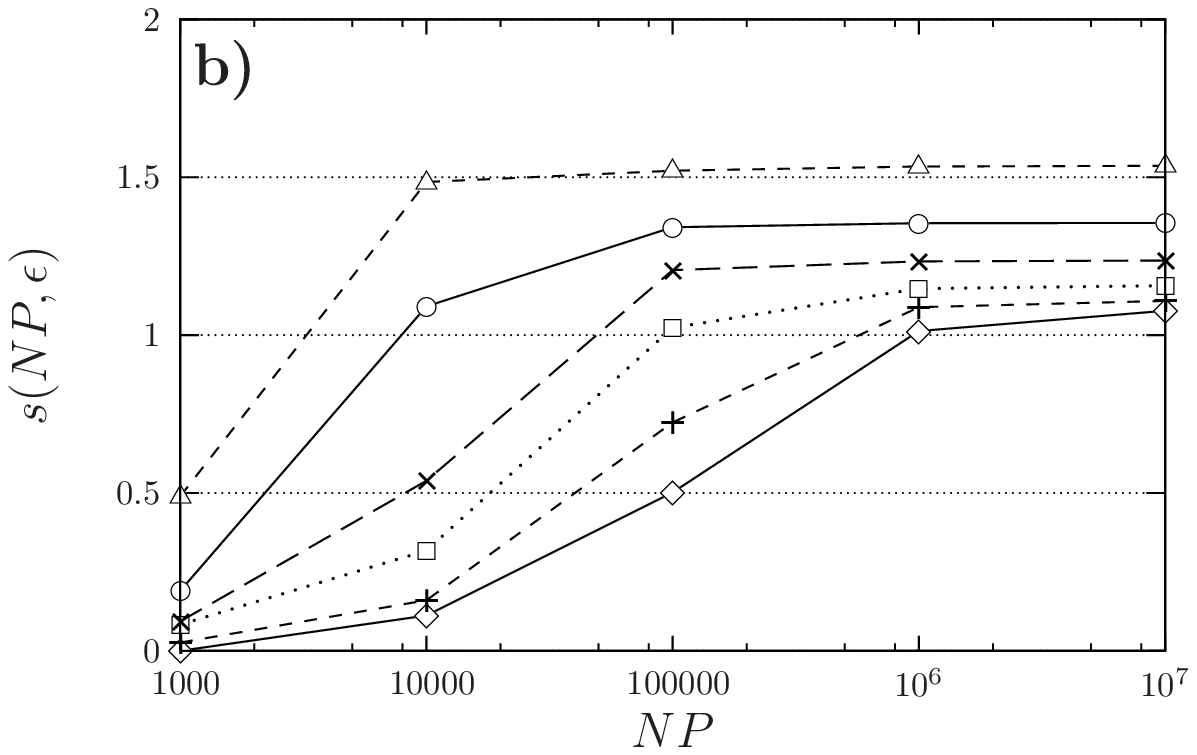}
	\caption{Same as fig.~\ref{fig6} but for $U=0.1$ (HMF) {\bf a)} and
	$K=0.2$ (standard map) {\bf b)}.}
	\label{fig7}
\end{figure}

Having understood that the limited number of particles is crucial in determining
the results, we show in fig.~\ref{fig6} and fig.~\ref{fig7} the dependence of 
the local slopes on the number of particles for both the HMF model and the standard map.
In all cases, $s$ is an increasing function of the number of particles, and,
at fixed number of particles, as the resolution increases, the slope decreases.
However, for a more chaotic phase space (fig.~\ref{fig6}), i.e. HMF at
$U=0.69$ and standard map at $K=1.2$, the curves corresponding to the higher resolution
show only a slight increase from zero slope and one cannot definitely conclude that,
as the number of particle increases, they will tend to slope 1 (while those
which converge, clearly point to dimension 2). On the contrary, for a less chaotic
phase space (HMF at $U=0.1$ and standard map at $K=0.2$), the curves corresponding to
the higher resolutions converge to slope (dimension) 1 as the number of particles 
increases, approximately to $(1/\epsilon)^2$.\footnote{For the standard map we have performed experiments for up
to $10^8$ particles. We have obtained a more evident levelling off 
of the slope as the number of particles increases and a convergence
to 1 from above as the resolution subsequently increases. For the HMF model the
convergence to $1$ is followed by a further drop to smaller values at higher
resolutions, as also shown in fig.~\ref{fig5}a: this is again a finite $N$
effect.}

We think that we have solved the puzzle of the existence of a fractal set 
generated in a finite time when starting with a one-dimensional set (a line)
in the phase space of Hamiltonian systems. {\it The fractal is not 
there and is a feature of the resolution with which the set is observed}.

Of course this does not diminish the role played by fractal concepts in systems with
long-range interactions. Certainly, when the initial set is already a fractal~\cite{maeda},
it's well possible that the set remains fractal during time evolution. Moreover,
when studying evolving universe models, one effectively introduces dissipative
terms in the equations of motion~\cite{maeda,miller} and one can indeed observe fractal sets generated
from non fractals. Another situation is the one of the advection of passive 
tracers~\cite{tel} or vector fields~\cite{ott}, where one indeed can observe
fractal sets in Hamiltonian phase spaces.  This is due to the presence
of non escaping orbits inside the mixing region, which form a zero measure
fractal repeller~\cite{tel} or, in the kinematic dynamo problem~\cite{ott},
to the concentration of the magnetic flux onto a singular set as $t\to\infty$.

\begin{ack}
We warmly thank F. Bagnoli for useful discussions and help in
numerical calculations. This work has been finantially supported by The
University of Florence, INFN, MIUR-COFIN03 contract {\it Order and chaos in
nonlinear extended systems} and the FIRB n. RBNE01CW3M\_01 project 
on synchronization.
\end{ack}


\begin{thebibliography}{99}

\bibitem{hiroko}
H. Koyama and T. Konishi, Phys.~Lett. A \textbf{279}, 226 (2001).

\bibitem{tsallis} V. Latora, A. Rapisarda, C. Tsallis, Phys. Rev. E
  \textbf{64}, 056134 (2001) and Physica A \textbf{305}, 129 (2002).
  
\bibitem{falconer}  K. Falconer, {\it Fractal geometry: mathematical foundations
and applications}, Wiley (1990).

\bibitem{hohlfeix} F. Hohl and M.R. Feix, Astroph. J., \textbf{147}, 1164 (1967).

\bibitem{antoniruffo} M. Antoni and S. Ruffo, Phys. Rev. E
\textbf{52}, 2361 (1995).
  
\bibitem{Spohn} H. Spohn, {\em Large Scale Dynamics of Interacting
    Particles}, Springer (1991).
    
\bibitem{chirikov} B.V. Chirikov, Phys. Rep.,\textbf{52}, 263 (1979).

\bibitem{HMFSpringer} T. Dauxois, V. Latora, A. Rapisarda, S. Ruffo,
 A. Torcini, {\em The Hamiltonian Mean Field Model: from Dynamics to
 Statistical Mechanics and back}, in {\it Dynamics and thermodynamics
 of systems with long range interactions}, T. Dauxois et al. Eds., Lecture
 Notes in Physics {\bf 602}, Springer (2002).
 
\bibitem{Yama} Y.Y. Yamaguchi, J. Barr\'e, F. Bouchet, T. Dauxois
and S. Ruffo, Physica A, {\bf 337}, 36 (2004).

\bibitem{ottino} J.M. Ottino, {\em The kinematics of mixing: stretching,
chaos and transport}, Cambridge University Press (1989).

\bibitem{benettin} G. Benettin, D. Casati, L. Galgani and A. Giorgilli,
Phys. Lett. A, \textbf{118}, 325 (1986).

\bibitem{maeda}
T. Tatekawa and K. Maeda, Astroph. J. \textbf{547}, 531 (2001).

\bibitem{miller}
B.N. Miller and J.L. Rouet, Phys. Rev. E \textbf{65}, 056121 (2001).

\bibitem{tel} T. Tel et al., Chaos, \textbf{10}, 89 (2000).

\bibitem{ott} E. Ott and T.M. Antonsen Jr., Phys. Rev. A, \textbf{39}, 3660 (1989).

\end{thebibliography}
\end{document}